\documentclass[amsmath,amssymb,aps,showkeys,floatfix,prd,a4paper,nofootinbib,onecolumn,notitlepage]{revtex4-1}

\usepackage{graphicx}
\usepackage{amsmath}
\usepackage{epstopdf}
\usepackage{amsfonts,amssymb}
\usepackage{float}
\usepackage[utf8]{inputenc}
\usepackage{pstricks}
\usepackage{color}
\usepackage[compat=1.0.0]{tikz-feynman}
\usepackage{simplewick}
\usepackage{colortbl}
\definecolor{Gray}{gray}{0.9}
\usepackage{float}
\usepackage{feynmp}
\usepackage{forest}
\usepackage{multirow}
\usepackage{verbatim}
\usepackage{hyperref}
\usepackage{tabularx}
\usepackage{bm}
\usepackage{subfigure}
\usepackage{verbatim}    
\usepackage{dcolumn}
\usepackage{bm}
\usepackage{epsfig}
\usepackage{braket}
\usepackage[normalem]{ulem} 
\usepackage{longtable,booktabs}
\usepackage{tikz}
\usetikzlibrary{tikzmark}
\usepackage[normalem]{ulem}

\newcommand{\be}{\begin{equation}}
\newcommand{\ee}{\end{equation}}
\newcommand{\ben}{\begin{eqnarray}}
\newcommand{\een}{\end{eqnarray}}

\def\MeV{\mbox{ MeV}}

\newcommand{\PreserveBackslash}[1]{\let\temp=\\#1\let\\=\temp}
\newcolumntype{C}[1]{>{\PreserveBackslash\centering}p{#1}}
\newcolumntype{R}[1]{>{\PreserveBackslash\raggedleft}p{#1}}
\newcolumntype{L}[1]{>{\PreserveBackslash\raggedright}p{#1}}

\begin{document}

\title{A note on the tensor and vector exchange contributions to $K \bar K \to K \bar K, D \bar D \to D \bar D $ and $\pi^+ \pi^-\to \pi^+ \pi^-$ reactions}

\date{\today}

\author{Luciano M. Abreu}
\email[]{luciano.abreu@ufba.br}
\affiliation{Instituto de Física, Universidade Federal da Bahia, Campus Ondina, Salvador, Bahia 40170-115, Brazil}
\affiliation{Instituto de F\'{\i}sica, Universidade de S\~{a}o Paulo, 
Rua do Mat\~{a}o, São Paulo SP, 05508-090, Brazil}

\author{Jing Song}
\email[]{Song-Jing@buaa.edu.cn}
\affiliation{School of Physics, Beihang University, Beijing, 102206, China}
\affiliation{Departamento de Física Teórica and IFIC, Centro Mixto Universidad de Valencia-CSIC Institutos de Investigación de Paterna, 46071 Valencia, Spain}

\author{Pedro C. S. Brandão}
\email[]{pedro.brandao@ufba.br}
\affiliation{Instituto de Física, Universidade Federal da Bahia, Campus Ondina, Salvador, Bahia 40170-115, Brazil}
\affiliation{Departamento de Física Teórica and IFIC, Centro Mixto Universidad de Valencia-CSIC Institutos de Investigación de Paterna, 46071 Valencia, Spain}

\author{ Eulogio Oset}
\email[]{oset@ific.uv.es}
\affiliation{Departamento de Física Teórica and IFIC, Centro Mixto Universidad de Valencia-CSIC Institutos de Investigación de Paterna, 46071 Valencia, Spain}


\begin{abstract}

In this note we study the tensor and vector exchange contributions to the elastic reactions involving the pseudoscalars mesons $\pi^+ \pi^-$, $K^{+}K^{-}$ and $D^{+}D^{-}$. In the case of the tensor-exchange contributions we assume that an intermediate tensor $f_2(1270)$ is dynamically generated from the interaction of two virtual $\rho$ mesons, with the use of a pole approximation. The calculation of the two-loop amplitude is facilitated since the triangle loops can be factorized and computed separately. The results show very small contributions coming from the tensor-exchange mechanisms when compared with those from the vector-exchange processes. {We compare our results for $\pi \pi$ and $K\bar{K}$ scattering
with those obtained in other works where the $f_2(1270)$ is considered as an ordinary $q\bar{q}$ meson. Our picture provides a smaller contribution but of similar order of magnitude for pion scattering and stabilizes the results in the case of $K\bar{K}$,  allowing us to make estimates for $D\bar{D}$ scattering.}

\end{abstract}

\maketitle

\section{\label{sec:level1}Introduction}

Despite the incredible knowledge on meson-meson interactions achieved thanks to the large amount of efforts and studies performed over the last five decades, this topic amazingly remains attracting attention of the particle physics community and being subject of intense debate. 
In a broad context, it is due to the fact that it provides detailed information on the hadronic properties and the hadronic mass spectra, which in its most profound sense yields a better comprehension of the strong interaction and the confinement mechanism. However, in general these analyses might not be a direct and simple task, because most of the mesons are resonances and not stable $q \bar q$  states. Strictly speaking, several mesons should not be interpreted as bound states obtained from a static potential in effective quark models of QCD, but as dynamically generated, i.e. stemming from the interaction of more elementary particles. Actually, a large number of meson states are broad resonances. We do not intend to present a more detailed discussion on this issue; for reviews we refer the reader to Refs.~\cite{Chen:2016qju,Hosaka:2016pey,Chen:2016spr,Lebed:2016hpi,Esposito:2016noz,Oset:2016lyh,Guo:2017jvc,Ali:2017jda,Olsen:2017bmm,Karliner:2017qhf,Yuan:2018inv,Liu:2019zoy,vanBeveren:2020eis}
 (reviews on hadron resonances within the lattice approach can also be found in Refs.~\cite{Dudek:2020aaf,Bulava:2022ovd,Mai:2022eur,Prelovsek:2023sta}).

We just introduce here some interesting examples: the broad scalar-isoscalar and scalar--isospin-$1/2$ resonances $\sigma / f_0(500)$ and $\kappa / K_0^*(700)$ have been subject of controversy over the years, but now they appear as established states in RPP 2023~\cite{Workman:2022ynf}. Their existence and properties have been extensively discussed in the reviews~\cite{Pelaez:2015qba,Pelaez:2021dak}, with a pedagogical analysis concerning the strong and weak points of the dispersive methods versus unitarized Chiral Perturbation Theory (UChPT) approaches~\cite{Oller1997,Kaiser1998,Dobado:1996ps,Oller:1998hw,Markushin:2000fa}. The resulting picture is that both $\sigma / f_0(500)$ and $\kappa / K_0^*(700)$ resonances are not ordinary quark-antiquark states, but their formation dynamics is dominated by pseudoscalar meson--pseudoscalar meson interactions, respectively the elastic $ \pi \pi $ scattering and $K \pi $ scattering. 
Another couple of interesting resonances in the light sector are the scalar-isoscalar $f_0(980)$ and scalar-isovector $a_0(980)$ resonances. They can be dynamically generated from the unitary pseudocalar--pseudoscalar coupled-channel chiral $\rm SU(3)$  approach, considering the channels $ \pi \pi ,  K \bar K, \eta \eta $ for $I=0$ and $ \pi \eta , K \bar K  $ for $I=1$~\cite{Oller1997,Liang:2014tia,Xie:2014tma,Abreu:2023hts}. In the tensor sector,  the tensor-isoscalar resonance $f_2(1270)$ emerges as a state generated from the vector meson-vector meson interaction, with the $\rho\rho$ channel being the dominant contribution~\cite{Geng2008,Molina:2008jw}. Interestingly, Refs.~\cite{Geng2008,Molina:2008jw} have made use of the so-called local hidden gauge formalism, which provides pseudoscalar-pseudoscalar-vector ($PPV$), vector-vector-vector ($VVV$) and four-vector ($VVVV$) structures~\cite{Bando:1987br}.

The picture of the $f_2(1270)$ as dynamically generated from the $\rho\rho$ is not free of controversy. Indeed, the $\rho\rho$ interaction in~\cite{Molina:2008jw,Geng2008} is obtained from $\rho$-exchange using the local hidden gauge approach~\cite{Bando:1987br,hidden2,hidden4,hideko}, but in the vector exchange propagator, $(q^{2} - m^{2}_{\rho})^{-1}$, the $q^{2}$ term is neglected, leading then to a contact interaction. This is a common practice in such calculations, since it is well known that with this procedure one obtains the chiral Lagrangians to the lowest order. An explicit derivation can be seen in Appendix $A$ of~\cite{diastoledo}. The equivalence of the local hidden gauge approach and chiral Lagrangians at higher order is also established, assuming vector meson dominance, in the work of~\cite{Ecker:1989yg}. Yet, if one sticks to $\rho$-exchange with the full structure of the $\rho$ propagation and projects over $S$-wave, one finds a singularity of this potential at $s = 3M^{2}_{\rho}$, when the exchanged $\rho$ meson can be placed on shell. This was shown in Ref.~\cite{Gulmez:2016scm}. This energy corresponds to $1334$ MeV, above the $f_2(1270)$ mass, which invalidates any conclusion using the on shell factorization of the Bethe-Salpeter equation, $T = [1-VG]^{-1}V$, at the singular point and any energy below it, as discussed in~\cite{Geng:2016pmf,Du:2018gyn}. To solve this problem a method was suggested in Section $4$ of~\cite{Gulmez:2016scm}, using a dispersion relation, which was further elaborated in~\cite{Du:2018gyn}. Yet, the method did not come to solve the problem, since as shown in~\cite{Molina:2019rai}, the dispersion relation method also had problems of convergence at the singular point and below. A solution to this issue was reported in~\cite{Geng:2016pmf}, evaluating exactly the loop functions which appear in the Bethe-Salpeter equation, which contain three and four propagators. This avoids the problem of the factorization of the potential in the evaluation of the loops and allows for a proper handling of these loops when the intermediate $\rho$ meson can be placed on shell. Surprisingly, the method gave rise to the $f_2(1270)$ as a bound $\rho\rho$ state and the couplings of the resonance  to the $\rho\rho$ channel were also in remarkable agreement with the simplified method of~\cite{Molina:2008jw,Geng2008} using a constant potential. The interpretation of the $f_2(1270)$ as dynamically generated from the $\rho\rho$ interaction is reinforced by the prediction of this picture in other processes. Indeed, in~\cite{Yamagata} it was shown that using the coupling from the $\rho\rho$ picture and vector meson dominance, a very good agreement was found for the decay widths of $f_2(1270)\to\gamma\gamma$ and $f_0(1370)\to\gamma\gamma$. The consistency of this interpretation has also been tested in radiative decays of $J/\psi$, $\psi(nS)$ and $\Upsilon(nS)$~\cite{albergeng,osethanhart,daioset,daixie}.
One can have a look at the problem from a different perspective. The $\rho\rho$ interaction in $J=2$ is attractive and much stronger than for $J=0$. The $J=0$ state $f_0(1370)$ is accepted as a $\rho\rho$ bound state in~\cite{Molina:2008jw,Geng2008,Gulmez:2016scm,Du:2018gyn}~\footnote{The $f_0$~(1370) appears as a $\rho\rho$ bound state at 1512 MeV in~\cite{Geng2008}, at $1532~\mathrm{MeV}$ in~\cite{Molina:2008jw}, in the range of 1476-1522 MeV in{~\cite{Gulmez:2016scm}} and in the range of 1410 - 1500~MeV 
in{~\cite{Du:2018gyn}}. In the PDG it appears in the bracket $1250-1440~\mathrm{MeV}$.}, 
hence a fortiori the $J=2$ state should be more bound. Independent of which is the technical mechanism by which the binding of this state could be found, we can use the coupling that emerges from assuming that one has a bound $\rho\rho$ state at $1270$ MeV. This coupling is given in~\cite{Gamermann:2009uq} as $g^{FT} = E(16\pi\gamma/\mu)^{1/2}$, derived in the limit of small binding,  with $E$ being the mass of the state, $\gamma = \sqrt{2\mu{B}}$, $\mu$ the $\rho\rho$ reduced mass and $B$ the biding energy of the state. The value of $g^{FT}$ for the $f_2(1270)$ assumed to be a bound $\rho\rho$ state using this formula is $g^{FT} = 10274$ MeV, which is in agreement with the value $10889$ MeV obtained in Ref.~\cite{Geng2008} or $10551$ MeV obtained in Ref.~\cite{Molina:2008jw} or those around $10000 - 11700$ MeV obtained in~\cite{Geng:2016pmf} with the improved method, where the discussion is made about the stability of this coupling in different approaches.
 That this formula, derived for the case of small binding, gives a value of the coupling so close to the one obtained with other methods, could be a coincidence. We have checked the formula in many examples of large binding with the results of the actual calculation, and one always finds similar results but not so close by.
 We can accept the range of values 1000-11700 of~\cite{Geng:2016pmf}, introducing uncertainties in the
final results which are not relevant when we conclude that the ratio of tensor to vector exchange in of the order of $10^{-3}-10^{-4}$.
This fact is important to note, because this coupling is the only thing we need in the derivation of the tensor exchange contribution.

For our present purposes, the relevant point in the hidden gauge formalism is that the $PPV$ vertex enables the analysis at tree-level of pseudoscalar meson--pseudoscalar meson interactions with an intermediate vector-exchange. For instance, one can evaluate the $\rho-$meson exchange mechanism contributing to the elastic scatterings $ \pi \pi ,  K \bar K $ mentioned above. Notwithstanding, one question appears in this scenario: {\it what about the contribution coming from tensor-exchange mechanism for the scattering matrix?} The tensor-exchange contribution can be evaluated for the case where the tensor is dynamically generated by the interaction of vector mesons as is the case in Refs.~\cite{Molina:2008jw,Geng2008,Geng:2016pmf} mentioned above. In this sense, another natural question is related to the relevance of the tensor-exchange mechanism with respect to the vector one.{This exercise was early addressed in Ref.~\cite{ARGYRES1974283} in the context of the $\pi N \to \pi N$ amplitude but there are many recent calculations based on chiral dynamics to which we turn below.}

The tensor exchange contribution was evaluated in Refs.~\cite{Donoghue:1988ed,Dobado:2001rv,Suzuki:1993zs,Katz:2005ir,Toublan:1995bk,Ananthanarayan:1998hj} looking at its contribution to the low energy constants (LECs) $l_{i}$'s of chiral perturbation theory within a SU$(2)$ formalism. Although there are differences in the results obtained, all of them give small contributions compared to the one of vector exchange to the same coefficient, which appears in the $\mathcal{L}_{4}$ Lagrangians, which is also small compared to the contribution of vector exchange in the $\mathcal{L}_{2}$ Lagrangian, the leading term in the interaction. A thorough discussion of these result, with an evaluation of the contribution to the related LECs in SU$(3)$, $L_{i}$'s, is done in~\cite{Ecker:2007us}. The result of this work is that the tensor exchange contributes to the $L_{3}$ coefficient, by an amount of $L^{(T)}_{3} = 0.16\times10^{-3}$, which is about $5\%$ of the value of $L_{3}$ from other sources, basically vector exchange, of $L_{3}\approx-3\times10^{-3}$. At this point it is worth noting how the derivation of Ref.~\cite{Ecker:2007us} is done. The input is based on the coupling of the $f_{2}(1270)$ to $\pi\pi$ and the $f_{2}(1270)$ width. In addition, the $f_{2}(1270)$ is assumed to be a $q\bar{q}$ state with an ideal mixture of the SU$(3)$ singlet and octet isosinglet states. In order to estimate the relevance of the $L^{(T)}_{3}$ term we should then see which is the contribution to $\pi\pi$ scattering. This is easily done by looking at the lowest order $T^{(I=0)}_{2}$ and $T^{(I=0)}_{4}$ amplitudes, the latter one evaluated by means of the $L^{(T)}_{3}$ coefficient, which can be looked in Eqs. (B3) and (B4) of Ref.~\cite{Oller:1998hw}. Choosing the threshold energy for reference we find

\begin{eqnarray}
    T^{(I=0)}_{2} & = & \frac{-7}{2f^{2}_{\pi}}m^{2}_{\pi}, \\
    T^{(I=0,T)}_{4} & = & -\frac{40}{f^{4}_{\pi}}L^{(T)}_{3}m^{4}_{\pi},
   \label{T2T4} 
\end{eqnarray}
{and then}
\begin{equation}
    R = \frac{ T^{(I=0,T)}_{4}}{T^{(I=0)}_{2}} = \frac{80}{7f^{2}_{\pi}}L^{(T)}_{3}m^{2}_{\pi} = 4\times10^{-3}.
\label{T2T4ratio}
\end{equation}

{This is a very small magnitude which allows the authors of Ref.~\cite{Ecker:2007us} to conclude that the tensor contribution is completely negligible. One may wonder whether this conclusion would be valid for $K\bar{K}$ scattering. In fact, it is easy to repeat the calculations done for $K\bar{K}$ scattering, using again the amplitudes of Ref.~\cite{Oller:1998hw} and, since the ratio of Eq. (\ref{T2T4ratio}) goes now as $m^{2}_{K}$, the ratio $R$ assumes values around $7\times10^{-2}$ $(R= 16L^{(T)}_{3}m^{2}_{K}/f^{2}_{K})$, still small but about $17$ times bigger than for $\pi\pi$ scattering. After this observation one might rightly wonder whether tensor exchange from $D\bar{D}$ scattering might be relevant. 
Certainly, the work of~\cite{Ecker:2007us} relying upon chiral symmetry and the Goldstone boson character of $\pi$ and $K$, cannot be extrapolated to the case of $D$ mesons, but our work, relying upon extrapolations of the local hidden gauge approach~\cite{Bando:1987br,hidden2,hidden4,hideko}, allows us to make estimates in this case too.
At this point it is worth mentioning that the work of Ref.~\cite{Ecker:2007us} assumes SU$(3)$ symmetry in the Lagrangians, which is well justified. However, it is also well known that, as soon as loops are considered with the physical masses of the particles, this symmetry is broken and this affects anything to do with dynamically generated states. An example is the $\Lambda(1405)$ and $\Lambda(1670)$ states. These states emerge from the interaction of $\bar{K}N, \pi\Sigma$ and other coupled channels~\cite{Kaiser:1995eg,Oset:1997it,Oller:2000fj,Jido:2003cb}, and it is shown in~\cite{Jido:2003cb} that if we assume SU$(3)$ symmetry with average masses of particles in the same SU$(3)$ multiplets, the two states are degenerate. When the physical masses are taken into account the degeneracy is broken and there is a splitting of the states by $250$ MeV. From this perspective it is important to investigate which are the consequences in the tensor exchange from the perspective of the $f_{2}(1270)$ being a dynamically generated state, a problem that we address in the present work.}

After the previous discussion, we believe that the evaluation and comparison of tensor and vector-exchange mechanisms deserve attention in the context of relevant meson-meson interactions treated with up-to-date frameworks. 
Thus, the purpose of this work is to analyze the tensor and vector ($\rho$) contribution mechanisms to the elastic reactions involving the pseudoscalars mesons $\pi^+\pi^-$, $K^{+}K^{-}$ and $D^{+}D^{-}$. The case of charmed mesons is also considered in order to explore the light-heavy sector and compare it to the light sector. In the case of the tensor-exchange contribution, we will follow Refs.~\cite{Geng2008,Molina:2008jw,Geng:2016pmf} and assume that the tensor $f_2(1270)$ is dynamically generated from an intermediate $\rho\rho$ interaction, which generates two-loop amplitudes to be calculated.

\section{Formalism}

\subsection{Effective formalism}

\begin{figure}[H]
    \centering
    \includegraphics[width=0.3\textwidth]{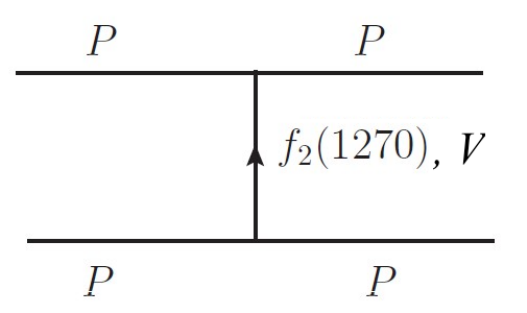}
    \caption{An example of one-meson exchange diagram for $ P P \to P P $ reactions.}
    \label{diaf2}
\end{figure}

\begin{figure}[H]
    \centering
\includegraphics[width=0.2\textwidth]{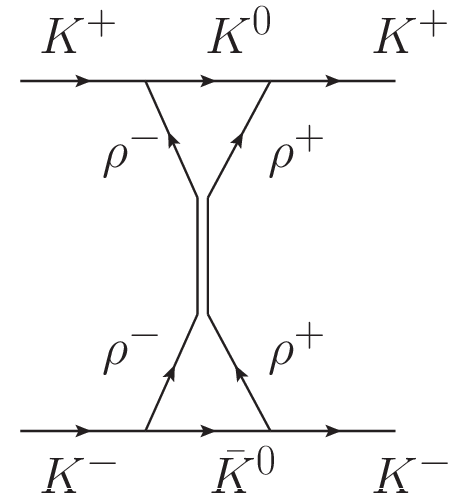}~~
\includegraphics[width=0.2\textwidth]{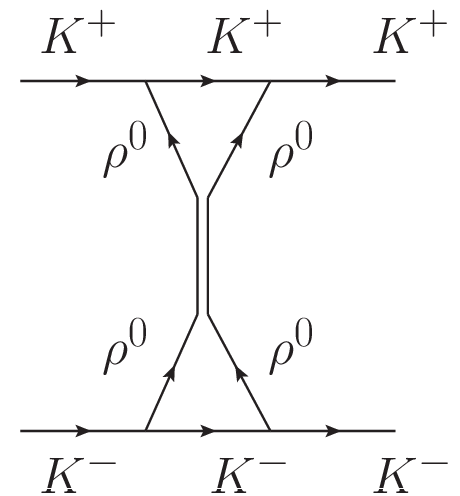}
    \caption{Tensor exchange diagrams for the $K^+ K^-\to K^+ K^-$ reaction. The tensor $f_2(1270)$ meson (represented by the double line) is dynamically generated from the interaction of the two intermediate $\rho$ mesons. Two more diagrams are present with $\rho^{0} \rho^{0},\rho^{-} \rho^{+}$ in the lower vertices of the first and second diagrams, respectively.}
    \label{Tv_K}
\end{figure}

\begin{figure}[H]
    \centering
\includegraphics[width=0.2\textwidth]{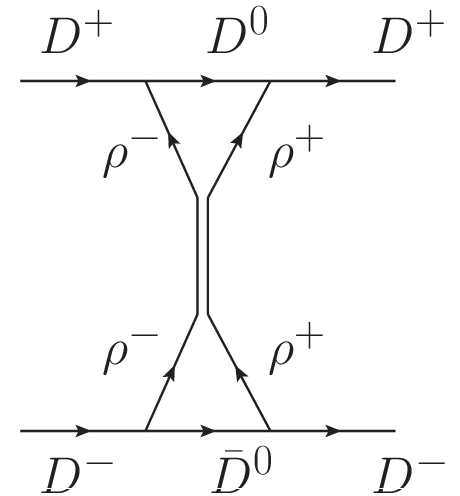}~~
\includegraphics[width=0.2\textwidth]{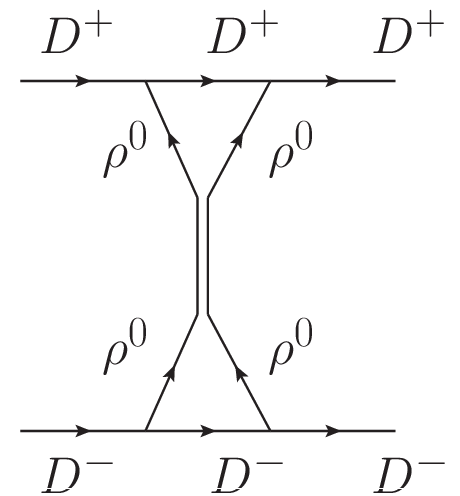}
    \caption{The same as in Fig.~\ref{Tv_K}, but for the $D^+ D^-\to D^+ D^-$ reaction. Two more diagrams are present with $\rho^{0} \rho^{0},\rho^{-} \rho^{+}$ in the lower vertices of the first and second diagrams, respectively. }
    \label{Tv_D}
\end{figure}

\begin{figure}[H]
    \centering
\includegraphics[width=0.2\textwidth]{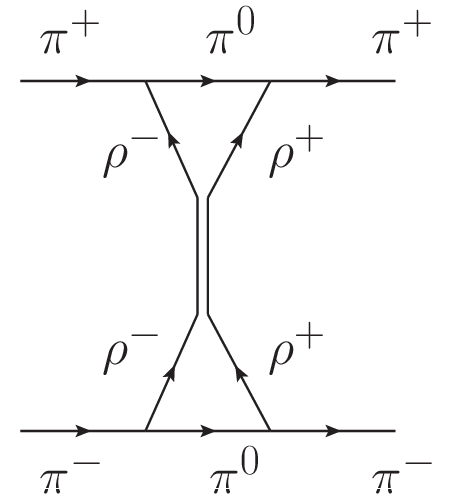}~~
\includegraphics[width=0.2\textwidth]{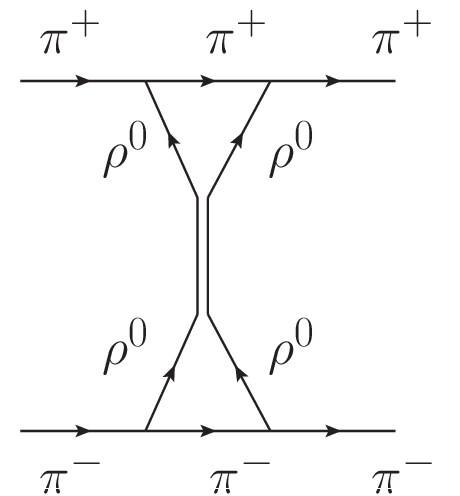}
    \caption{The same as in Fig.~\ref{Tv_K}, but for the $\pi^+ \pi^-\to \pi^+ \pi^-$ reaction. Two more diagrams are present with $\rho^{0} \rho^{0},\rho^{-} \rho^{+}$ in the lower vertices of the first and second diagrams, respectively. }
    \label{Tv_PI}
\end{figure}

We start by presenting the mechanism for the $PP\rightarrow PP$ reaction via one-meson exchange, depicted in Fig.~\ref{diaf2}, usually studied considering the $\rho$ meson as the intermediate particle. Strictly speaking, its amplitude can be directly computed employing an effective $PPV$ coupling. However, our interest here is also to evaluate the tensor-exchange contribution. In this sense, as a first attempt let us focus on the lightest spin-two resonance, i.e. the $f_2(1270)$ state, {the one considered in Ref.~\cite{Ecker:2007us}.} Refs.~\cite{Molina:2008jw,Geng2008,Geng:2016pmf} pointed out that it can be interpreted as a dynamically generated resonance mostly from the $\rho\rho$ interaction in the $S-$wave $I=0, J=2$ channel. Thus, we benefit from these preceding studies and consider the intermediate $f_2(1270)$ formed from the two $\rho-$mesons. From the former perspective, the mechanisms for the reactions we pay attention are shown in Figs.~\ref{Tv_K},~\ref{Tv_D} and~\ref{Tv_PI}, respectively in the $K^+ K^-\to K^+ K^-$, $D^+ D^-\to D^+ D^-$ and $\pi^+ \pi^-\to \pi^+ \pi^-$ processes. We estimate them at the respective thresholds. 
 

Let us first determine the amplitude for $K^+ K^-\to K^+ K^-$ reaction. Fig.~\ref{Tv_K} shows the two possible contributions coming from distinct intermediate $\rho \rho K$ triangles, i.e. $\rho^{-} \rho^{+} K^0 (\bar{K}^{0})$ and $\rho^{0} \rho^{0} K^+ (K^{-})$. The couplings $\rho KK$ can be treated effectively by using the following structure~\cite{Molina:2008jw,Geng2008,Bando:1987br}, 
\begin{align}\label{LPPV}
    \mathcal{L}_{PPV}= -ig \langle [P,\partial_\mu P] V^\mu\rangle,
\end{align}
where $ g $ is the coupling constant $g=m_V/(2f_\pi)$ ($m_V=800$ MeV, $f_\pi=93$ MeV); $ \left\langle\ \cdots \right\rangle $ is the trace over the flavor space; and $P,V$ denote the  $ q \bar{q} $ matrices in $ \text{SU}(4) $ flavor space written in terms of pseudoscalar or vector mesons:
\begin{align}\label{meson_P}
\centering
P=
\left(
  \begin{array}{cccc}
    \frac{\pi^0}{\sqrt{2}}+\frac{\eta}{\sqrt{3}}+\frac{\eta'}{\sqrt{6}} & \pi^{+} & K^{+} & \Bar{D}^{0}\\
    \pi^{-} & \frac{-\pi^0}{\sqrt{2}}+\frac{\eta}{\sqrt{3}}+\frac{\eta'}{\sqrt{6}} & K^0 & D^{-}\\
    K^{-} & \bar{K}^0 & -\frac{\eta}{\sqrt{3}}+\sqrt{\frac{2}{3}}\eta' & D_{s}^{-} \\
    D^{0} & D^{+} & D_{s}^{+} & \eta_{c} \\
  \end{array}
\right),~~~~
\end{align}
\begin{align}\label{meson_V}
\centering
V_{\mu}=
\left(
  \begin{array}{cccc}
    \frac{\rho^0}{\sqrt{2}}+\frac{\omega}{\sqrt{2}} & \rho^{+} & K^{*+} & \Bar{D}^{*0} \\
    \rho^{-} & \frac{-\rho^0}{\sqrt{2}}+\frac{\omega}{\sqrt{2}} & K^{*0} & D^{*-} \\
    K^{*-} & \bar{K}^{*0} & \phi & D_{s}^{*-}\\
    {D}^{*0} & D^{*+} & D_{s}^{*+} & J/\Psi \\
  \end{array}
\right)_{\mu};
\end{align}
in the $P$ matrix we have taken the standard $\eta-\eta'$ mixing of Ref.~\cite{Bramon:1992kr}. Although the $ \text{SU}(4) $ formalism is used, one can see that the vertices can be evaluated explicitly employing wave functions in terms of quarks and the only elements of $ \text{SU}(4) $ needed are the structure of the mesons as $q \bar q$, or reciprocally the  $q_j \bar q_j$ components written in terms of physical mesons~\cite{Sakai:2017avl}. {Even then one would be making some SU$(4)$ symmetry assumption by assuming the same $g$ coupling in Eq.~(\ref{LPPV}) in all sectors. Yet, if one looks at the diagrams in Fig.~\ref{Tv_D} and the microscopic picture of Ref.~\cite{Sakai:2017avl}, one observes that the $\rho$ only couples to the light quarks of the $D$ mesons and the $c$ quarks are spectators. One is hence only using the SU$(3)$ subgroup of SU$(4)$, since the strange quarks do not come into play.}

The next step is to determine the $f_2(1270)-$exchange contributions depicted in Fig.~\ref{Tv_K}. As discussed above, the $S-$wave $I=0, J=2$ channel of the $\rho\rho$ interaction is the dominant mechanism for the formation of the $f_2(1270)$ resonance~\cite{Molina:2008jw,Geng2008,Geng:2016pmf}. 
By looking at the structure 
\begin{align}
\epsilon_{\mu}(\rho, 1) \epsilon_{\nu}(\rho, 2) \to \epsilon_{\mu'}(\rho, 3) \epsilon_{\nu'}(\rho, 4), 
\label{polvec}
\end{align}
the $ J=2 $ projected amplitude goes with the combination~\cite{Molina:2008jw,Geng2008} 
\begin{align}
P^{(2)} = \frac{1}{2} \left[ \epsilon_{\mu}(1) \epsilon_{\nu}(2)  \epsilon^{\mu}(3) \epsilon^{\nu}(4) + \epsilon_{\mu}(1) \epsilon_{\nu}(2)  \epsilon^{\nu}(3) \epsilon^{\mu}(4) \right] - \frac{1}{3} \epsilon_{\mu}(1) \epsilon^{\mu}(2)  \epsilon_{\nu}(3) \epsilon^{\nu}(4),
\label{projector}
\end{align}
and hence the propagator of the $f_2(1270)$ goes as 
\begin{align}
    D (k) = \frac{g_{f_2}^2}{k^2-m_{f_2}^2} P^{(2)},
\label{propf2}
\end{align}
where $g_{f_2}$ is the coupling whose value calculated in~\cite{Molina:2008jw} is $10551$ MeV, while in~\cite{Geng2008} it is $10889$ MeV. 



We remark that the $ |\rho\rho \rangle $ isospin state with $ I=0 $ adopted here is given by 
\begin{align}
    |\rho\rho, ~~I=0 \rangle \equiv -\frac{1}{\sqrt{6}}( \rho^+\rho^- + \rho^-\rho^+ + \rho^0\rho^0 ). 
\end{align}
The extra factor $1/\sqrt{2}$ is due to the unitary normalization employed for the counting of identical particles in the intermediate states.

\subsection{Calculation of the loop contribution}




\begin{figure}[H]
    \centering
\includegraphics[width=0.25\textwidth]{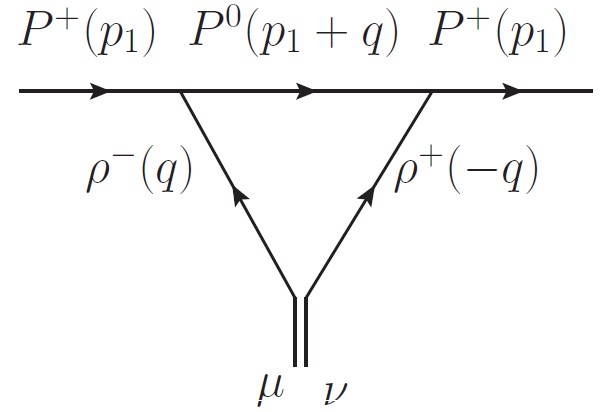}~~
\includegraphics[width=0.25\textwidth]{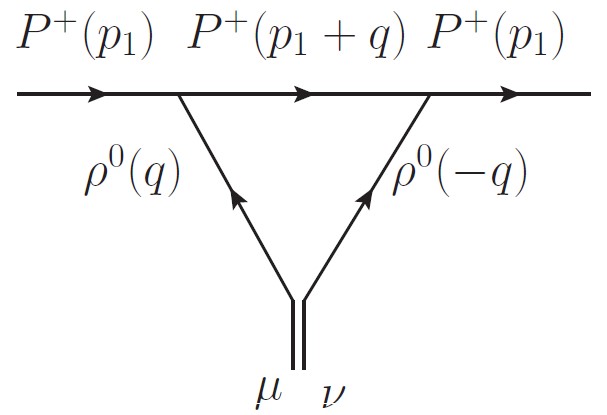}
    \caption{Upper vertices of the diagrams shown in Figs.~\ref{Tv_K}, \ref{Tv_D} and~\ref{Tv_PI}. $P (= K, \pi , D)$ denote the pseudoscalar meson in the reaction; the vertex in the left (right) is associated to the corresponding upper vertex in the left (right) diagrams of Figs.~\ref{Tv_K}-\ref{Tv_PI}. }
    \label{VMNP}
\end{figure}

The calculation of the two-loop diagrams in Figs.~\ref{Tv_K}-\ref{Tv_PI} is facilitated by the fact that the upper and lower vertices can be factorized due to the presence of the $f_2$-exchange amplitude written in Eq.~(\ref{propf2}). As a consequence, each loop can be computed separately, and the final amplitude will be the product of these two contributions by the $f_2$-exchange amplitude written in Eq.~(\ref{propf2}). Then, we pay attention to the upper vertices of the diagrams in Figs.~\ref{Tv_K}-\ref{Tv_PI} (denoted here by $V_{\mu \nu}$), and depict them in Fig.~\ref{VMNP}.
By using Eq.~(\ref{LPPV}) for the $ \rho P P  $ vertices of the type $\epsilon_{\alpha} (2 p_1 + q)^{\alpha}$, and summing over polarization of the vector mesons, 
\begin{align}
\sum _{pol} \epsilon_{\alpha} \epsilon_{\mu}  = - g_{\alpha  \mu} + \frac{k_{\alpha}  k_{\mu} }{m_{V}^2},
\label{trans_proj}
\end{align}
we obtain the analytical expression
\begin{eqnarray}
 -iV_{\mu \nu} & = & - \frac{3}{2} \frac{1}{\sqrt{6}}g_{f_2} g^2 \int \frac{d^4q}{(2\pi)^4}\left( \frac{1}{q^2-m_{\rho}^2+i\epsilon} \right)^2 \frac{1}{(p_1 + q)^2-m^2_{K}+i\epsilon }
\nonumber \\
& & \times \left[ -(2 p_1 + q )_{\mu} + \frac{1}{m_{\rho}^2} (2 p_1 + q )\cdot q  q_{\mu} \right] 
\left[ -(2 p_1 + q )_{\nu} +  \frac{1}{m_{\rho}^2} (2 p_1 + q )\cdot q  q_{\nu} \right] , 
 \label{vertex1}
\end{eqnarray}
We remark that the factor $\frac{3}{2}$ on the right hand side of this expression comes,  using the Lagrangian of{ Eq.~(\ref{LPPV})}, from the following fact: the calculation of the upper vertex on the right hand side of Fig.~\ref{Tv_K} can be done similarly, with the only difference appearing from the coupling involving the $\rho^0$ meson, which has an additional factor $1/\sqrt{2}$ with respect to that involving  $\rho^{\pm}$ meson. Therefore, we have added these two contributions in Eq.~(\ref{vertex1}) by taking this factor $1+(1/\sqrt{2})^2=3/2$, since in the end we are interested on the total amplitude of the reaction. 

For the case of $D\bar{D}$ scattering the loop function gives the same result except that the masses are different.
However, for the case of $\pi\pi$ scattering the use of { Eg.~(\ref{LPPV})} leads to a different coefficient, and instead of the factor $\frac{3}{2}$ we obtain a factor 4. This relative factor of $\left(\frac{3}{8}\right)$ from the pion case to that of the Kaon will be important to interpret the results that we obtain.

The integral in Eq.~(\ref{vertex1}) is linearly divergent in the $q^0$ integration due to the terms proportional to $\frac{1}{m_{\rho}^2}$. Notwithstanding, let us assume the presence of a form factor that makes the integration over $q^0$ convergent and therefore can be calculated via the Cauchy's residues (the actual regulator of the loop will be considered later). First, to make the integration easier,  we use the formula 
\begin{eqnarray}
\left( \frac{1}{q^2-m_{\rho}^2+i\epsilon} \right)^2  = \frac{\partial }{\partial m_{\rho}^2}  \left( \frac{1}{q^2-m_{\rho}^2+i\epsilon} \right), 
\label{vertex2}
\end{eqnarray}
and separate the positive and negative energy parts of the $\rho$ and $K$ propagators as
\begin{eqnarray}
\frac{1}{q^2-m^2+i\epsilon}=\frac{1}{2\omega_i }\left(\frac{1}{q^0-\omega_i +i\epsilon} 
                 - \frac{1}{q^0+\omega_i -i\epsilon} \right),
 \label{eq23}
\end{eqnarray}
with $\omega_i \equiv \omega_i(\vec q)=\sqrt{\vec q^2+m_i^2}$. Using the appropriate paths, the Cauchy integration over $q^0$ selects the poles $q^0 = \pm \omega_{\rho}$, and we obtain the loop amplitude
\begin{eqnarray}
 -iV_{\mu \nu} & = & -\frac{3}{2} \frac{1}{\sqrt{6}}g_{f_2} g^2 i \frac{\partial }{\partial m_{\rho}^2}  \int \frac{d^3 q}{(2\pi)^3} \frac{1}{2\omega_{\rho} \omega_{K}} \nonumber \\
 &  \times & \left\{ \frac{1}{p_1^0 + \omega_{\rho} + \omega_{K}-i\epsilon} \left[ \left( (2 p_1 + q )_{\mu} - \frac{1}{ m_{\rho}^2} (2 p_1 + q )\cdot q  q_{\mu} \right) 
\left( (2 p_1 + q )_{\nu} - \frac{1}{m_{\rho}^2} (2 p_1 + q )\cdot q  q_{\nu} \right) \right]_{q^0 = + \omega_{\rho}} \right.
\nonumber \\
 &  - & \left. \frac{1}{p_1^0 - \omega_{\rho} - \omega_{K}+i\epsilon} \left[ \left( (2 p_1 + q )_{\mu} - \frac{1}{ m_{\rho}^2} (2 p_1 + q )\cdot q  q_{\mu} \right) 
\left( (2 p_1 + q )_{\nu} -  \frac{1}{m_{\rho}^2} (2 p_1 + q )\cdot q  q_{\nu} \right) \right]_{q^0 = - \omega_{\rho}} \right\}. \nonumber \\
 \label{vertex3}
\end{eqnarray}

Recalling that we are working at the threshold of the pseudoscalar mesons, then the external tri-momentum is zero, and as a consequence the components $V_{0i}, V_{i0} \sim k_i = 0 $.  Moreover, the component $V_{ij}$ has the structure
\begin{align}
    V_{ij} = a\delta_{ij} + b k_{i}k_{j},
\end{align}
the $b$ term vanishes because $k_i = 0$, and the $\delta_{ij}$ in both vertices vanishes with the combination $[\frac{1}{2}(\epsilon_i\epsilon_j\epsilon_i\epsilon_j+\epsilon_i\epsilon_j\epsilon_j\epsilon_i)-\frac{1}{3}\epsilon_i\epsilon_i\epsilon_j\epsilon_j]$ of Eq.~(\ref{projector}).
Thus, only the $V_{00}$ component contributes. In the sequence, we note that for $\mu = \nu = 0 $ the quantities between brackets in the second and third lines of Eq.~(\ref{vertex3}) read
\begin{eqnarray}
 \left[ \left( (2 p_1 + q )_{\mu} - \frac{1}{m_{\rho}^2} (2 p_1 + q )\cdot q  q_{\mu} \right) 
\left( (2 p_1 + q )_{\nu} -  \frac{1}{m_{\rho}^2} (2 p_1 + q )\cdot q  q_{\nu} \right) \right]_{q^0 =  \pm \omega_{\rho} ; \mu = \nu = 0 } =  \left( \frac{2 E_1 \vec{q}^{\ 2}}{m_{\rho}^2} \right)^2 , 
\label{vertex4}
\end{eqnarray}
{with $E_{1}$ being the energy of the incoming meson ($m_{i}$ at threshold)}, which allows to write the $V_{00}$ component of $V_{\mu \nu}$ in Eq.~(\ref{vertex3}) in its final form, 
\begin{eqnarray}
 -iV_{ 00 } & = &   C_j \frac{1}{\sqrt{6}}g_{f_2} g^2 i \frac{\partial }{\partial m_{\rho}^2}  \int \frac{d^3 q}{(2\pi)^3} \frac{1}{2\omega_{\rho} \omega_{K}} 
 \frac{\omega_{\rho} + \omega_{K}}{E_1^2 -  (\omega_{\rho} + \omega_{K})^2 + i\epsilon}  \left( \frac{2 E_1 \vec{q}^{\ 2}}{m_{\rho}^2} \right)^2  \ \Theta(q_\mathrm{max} - |\vec{q}|)  \left(\frac{\Lambda^2}{\Lambda^2+\Vec{q}~^2}\right)^2, 
 \label{vertex5}
\end{eqnarray}
with $C_j=\frac{3}{2}$ for $j=K,~D$ and  $C_j=4$ for $j=\pi$,
where we have introduced a regulator with a cutoff $q_\mathrm{max} $ which is inherent to the $\rho \rho \to \rho \rho $ amplitude, since the chiral unitarity approach with the meson-meson--loop function regularized with $q_\mathrm{max} $ implies a $t-$matrix of the type $t(q,q') \Theta(q_\mathrm{max} - |\vec{q}|) \Theta(q_\mathrm{max} - |\vec{q}'|)$~\cite{Gamermann:2009uq}. The value of $q_\mathrm{max}$ is hence the same as the one used in~\cite{Molina:2008jw,Geng2008} to obtain the $VV \to VV$ amplitudes. In addition, we also include the form factor $(\frac{\Lambda^2}{\Lambda^2+\Vec{q}~^2})^2$ for each $\rho\pi\pi$ ($K\bar{K}, D\bar{D}$) vertex, which was used in~\cite{Molina:2008jw,Geng2008} in the box diagrams to calculate the width of the $f_2(1270)$. We take $\Lambda =1200$~MeV as in~\cite{Molina:2008jw}.

The calculation of the lower vertices in Fig.~\ref{Tv_K} is also performed using the same techniques summarized above, and we find that the $V_{00}$ component of the lower vertices are the same as for those of the upper vertices.  

Finally, we can combine all ingredients discussed before together with the structure of the $J=2$ projector of Eq.~(\ref{projector}) and write the total amplitude of the $K^+ K^-\to K^+ K^-$ reaction with tensor exchange shown in Fig.~\ref{Tv_K} as
\begin{eqnarray}
 -i T^{(T)} (k) & = &  \frac{i}{ k^2 - m_{f_2}^2 }\left\lbrace \frac{1}{2} \left[(-i)V_{ 00 }(-i)V_{ 00 } + (-i)V_{ 00 }(-i)V_{ 00 } \right] - \frac{1}{3}(-i)V_{ 00 }(-i)V_{ 00 } \right\rbrace  . 
 \label{Tamplitude1}
\end{eqnarray}
Hence, at threshold it becomes
\begin{eqnarray}
 T ^{(T)} (0) & = &  - \frac{2}{ 3 m_{f_2}^2 } \left[V_{ 00 }\right]^2 . 
 \label{Tamplitude2}
\end{eqnarray}
This contribution is negative, evidencing the attractive character of this process. 

{It is interesting to note that Eq.~(\ref{Tamplitude2}) together with Eq.~(\ref{vertex5}) tell us that the tensor contribution is proportional to $m^{4}_i$, as in Ref.~\cite{Ecker:2007us} (see Eq.~(\ref{T2T4})). It is also proportional to $g^{4} = (m_{V}/2f)^4 \sim f^{-4}$ as in Eq.~(\ref{T2T4}).}

In order to compare this tensor-exchange mechanism with the one from the exchange of a vector meson, we should calculate an equivalent diagram shown in Fig.~\ref{diaf2} but replacing $f_2$ by a $\rho^0$ meson. Then, the use of the local hidden gauge formalism outlined in Eqs.~(\ref{LPPV}) and (\ref{meson_V}) gives the following amplitude for a $\rho-$meson exchange for the cases of Kaon and $D$ meson:
\begin{eqnarray}
 -i T ^{(V)} (k) & = &  \frac{(-i g)}{\sqrt{2}}(2 p_1 + k )^{\mu} \frac{i}{ k^2 - m_{\rho}^2 } \left( - g_{\mu  \nu} + \frac{k_{\mu}  k_{\nu} }{m_{\rho}^2} \right) \frac{(i g)}{\sqrt{2}}(2 p_1 - k )^{\nu}.  
 \label{Tamplitude3}
\end{eqnarray}
 The case of the pion is characterized by an amplitude 4 times bigger than that given above. As a consequence, at threshold $(k=0)$ Eq.~(\ref{Tamplitude3}) reads
\begin{eqnarray}
  T^{(V)} (0)  =    
\begin{cases}
    - \frac{2 g^2 }{ m_{\rho}^2 }  E_1^2~ (\mathrm{for}~~K,~D);\\
    - \frac{8 g^2 }{ m_{\rho}^2 }  E_1^2~(\mathrm{for}~~\pi). 
\end{cases}
 \label{Tamplitude4}
\end{eqnarray}

A final comment on this section is that the computation of the amplitudes for the $D^+ D^-\to D^+ D^-$ and $\pi^+ \pi^-\to \pi^+ \pi^-$ processes depicted in Figs.~\ref{Tv_D} and~\ref{Tv_PI} are done analogously to the $K^+ K^-\to K^+ K^-$ case, just by using the corresponding masses of the respective pseudoscalar mesons. Thus, we will also utilize Eqs.~(\ref{Tamplitude3}) and~(\ref{Tamplitude4}) for these mentioned processes. 

\begin{figure}[H]
\centering
\subfigure[]{\includegraphics[width=0.40\textwidth]{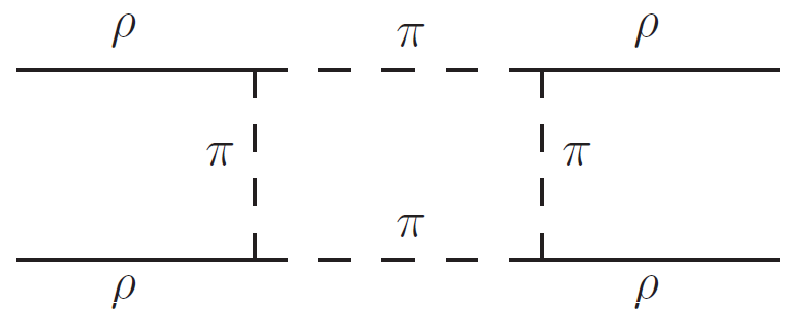} 
\label{box1}}  ~~
\put(0,40){\mbox{$\Longrightarrow$}} \hspace{0.4cm} ~~
\subfigure[]{\includegraphics[width=0.40\textwidth]{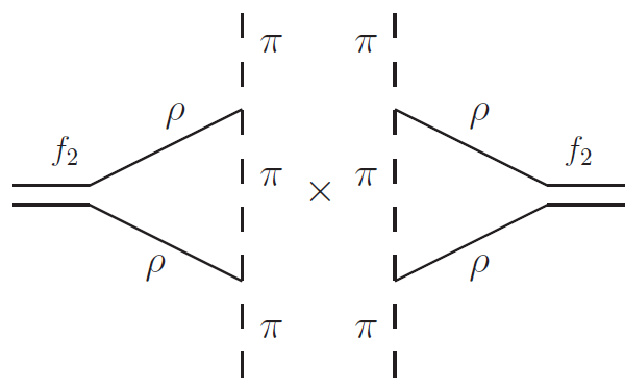}
\label{box2}}
\caption{(a) Box diagram considered in~\cite{Molina:2008jw,Geng2008} to account for the $f_{2}(1270)\to\pi\pi$ decay. (b) Break up of diagram (a) to show the connection with the diagrams of Fig.~\ref{Tv_PI}.}
\label{box}
\end{figure}

{At this point it is relevant to show that our mechanism contains the same ingredients than Ref.~\cite{Ecker:2007us}, based on the coupling of $f_{2}(1270)$ to $\pi\pi$. Indeed in Refs.~\cite{Molina:2008jw,Geng2008}, apart from the vector-vector as leading terms, box diagrams as shown in Fig.~\ref{box} were considered in~\cite{Molina:2008jw,Geng2008} and added to the leading terms to account for $f_{2}(1270)\to\pi\pi$, and a successful picture of this decay was obtained in~\cite{Molina:2008jw,Geng2008}. Note that the vertices shown explicitly in~Fig~\ref{box2} are exactly those in Fig.~\ref{Tv_PI}.}

{Thus, our picture is using an input consistent with the coupling of $f_{2}(1270)$ to $\pi\pi$. The difference with Ref.~\cite{Ecker:2007us} stems from the explicit picture of the mechanism used to couple $f_{2}(1270)$ to two pions. There, an assumption is made with the ideal mixture of singlet and octet $SU(3)$ isomultiplets coupling to two pions, while here $f_{2}(1270)$ is dynamically generated via $\rho\rho$ interaction, and the coupling to $\pi\pi$ proceeds via the triangle vertices of Fig.~\ref{Tv_PI}. Then, from construction in~\cite{Ecker:2007us}, and as a consequence of the dynamics in~\cite{Molina:2008jw,Geng2008}, the two pictures coincide when two pions are emitted from the $f_{2}(1270)$. However, when one has a incoming and an outgoing pion, as in the vertices of Fig.~\ref{Tv_PI}, the two pictures will be different. There is also an interesting consequence of using the present picture. While an implicit SU$(3)$ picture is assumed in Ref.~\cite{Ecker:2007us} to go from $\pi\pi$ to $K\bar{K}$ scattering, here we have loops, and in $\pi\pi$ scattering we have pions as intermediate states (Fig.~\ref{Tv_PI}), while in $K\bar{K}$ scattering we have Kaons (Fig.~\ref{Tv_K}), and in $D\bar{D}$ we have $D$ mesons (Fig.~\ref{Tv_D}), and the loop functions have different weights ($C_j$ coefficients).  We shall come back to this point in next section.} 


\section{Results}

Here the results are presented by computing the ratio between the contributions for the different processes coming from the tensor-exchange mechanism and the one with exchange of a vector meson, i.e.
\begin{eqnarray}
R = \left| \frac{T^{(T)} (0)}{T^{(V)} (0)} \right|, 
\label{ratio}
\end{eqnarray}
where the amplitudes $T^{(T)} (0)$ and $T^{(V)} (0)$ are given in Eqs.~(\ref{Tamplitude2}) and~(\ref{Tamplitude4}).

In Table \ref{Tab1} we show the ratio $R$ for the processes $\pi^+ \pi^-\to \pi^+ \pi^-$, $K^+ K^-\to K^+ K^-$ and $D^+ D^-\to D^+ D^-$ taking different values for the coupling $g_{f_2}$ calculated in Refs.~\cite{Geng2008} and~\cite{Molina:2008jw}, and for the cutoff $q_\mathrm{max}$. 

Firstly, we note that since the coupling $g_{f_2}$ from~\cite{Geng2008} is about $3 \%$ bigger than the one from~\cite{Molina:2008jw}, then the ratios calculated using the latter case are obviously greater than the former one by a factor of about  $6-7 \%$. Secondly, concerning the dependence with the cutoff, as expected the ratios suffer an augmentation as $q_\mathrm{max} $ increases. In particular, within the range of $q_\mathrm{max}$ considered, i.e. $600 \MeV \leq q_\mathrm{max} \leq  900 \MeV$, the ratios present a variation of about one order of magnitude.
\footnote{In Refs.~\cite{Geng2008,Molina:2008jw} the loops were regularized by means of dimensional regularization. The equivalent cutoff needed with the cutoff regularization were in that range of values.  }

However, the most important finding of this analysis is that the ratios are of the order of $10^{-2}$ or $10^{-4}$, showing that the contributions coming from the tensor-exchange mechanism are highly suppressed in comparison with those from the vector-exchange processes. One can think that this suppression might come mainly from the difference between the tensor and meson masses present in the intermediate propagators given in Eqs.~( \ref{Tamplitude2}) and~( \ref{Tamplitude4}). However, the ratio $m_{f_2}^{-2}/ m_{\rho}^{-2}$, gives a factor of about 0.4, which means that the small magnitude is mainly due to the vertex contribution $V_{ 00 }$. 

It is interesting to compare our result with those obtained from the use of $L^{(T)}_{3}$ of~\cite{Ecker:2007us}. 
Let us take $g_{f_2}=10551$~MeV and $q_\mathrm{max}=850$~MeV for comparison, the value taken in Ref.~\cite{Molina:2008jw}. For pion scattering our result is about a factor $2.6$ smaller than 
in~\cite{Ecker:2007us} (see Eq.~(\ref{T2T4ratio})), which can be understood as follows. In Refs.~\cite{Geng2008,Molina:2008jw,Geng:2016pmf}  the $f_2(1270)$ coupling to two pions is taken into account by adding the box diagram of Fig.~\ref{box}(a) to the $\rho\rho$ potential and solving
the Bethe–Salpeter equation. A soft form factor, $\Lambda^2/(\Lambda ^2+\Vec{q}~^2)$ at each $\rho \pi\pi$ vertex, is introduced in the box diagram and fine tuning is done to obtain the $f_2(1270)$ width, which according to the PDG~\cite{Workman:2022ynf} comes in a {proportion} of $84 \% $ from $\pi \pi$. The $\rho\rho$  decay channel, which accounts for  about $10 \% $
in the PDG is also accounted for in~\cite{Geng2008,Molina:2008jw} by considering the mass distribution of the $\rho$. The small $K\bar{K}$ channel is also considered in~\cite{Geng2008} and found of the order of $10 \%$ in  line with the $4.6 \%$ of the PDG. 
This means that our picture {produces}, with the small {fine tuning}, a good reproduction of $f_2(1270) \to \pi\pi $.

We can be more concrete. The choice of the effective potential and regulator $q_{max}$ guarantees that the $f_2(1270)$ is obtained with the right mass around $1270$ MeV. It changes around $1240-1280$ MeV~\cite{Molina:2008jw}, depending on the choice of $q_{max}$. The width of this state appears finite in the approach, but fine tuning with the form factor $\Lambda^2 / (\Lambda^2 + \vec{q}^2)$ gives rise in~\cite{Molina:2008jw} to the total width $ \Gamma \simeq 140$ MeV, compared to the experimental one $\Gamma^{(exp)} \simeq 186 $ MeV. 
Using the splitting of the total width into partial decay widths of~\cite{Geng2008}, we get $88\%$ for $\pi \pi $ decay, which means  $\Gamma_{\pi \pi} \simeq 123 $ MeV, compared with the experimental one of  $\Gamma_{\pi \pi}^{(exp)} \simeq 157 $ MeV. The partial decay width into $K \bar{K}$ is about $10\%$ of the total, which means  $ \Gamma_{K \bar{K}} \simeq 14 $ MeV, compared with the experimental one $\Gamma_{K \bar{K}} \simeq 8.6 $ MeV. Considering the differences between our calculated partial decay widths and the experimental ones, we can induce that we could be underestimating the rate of the tensor exchange in the case of pions by about $25\%$, and overestimating the rate for the case of kaons by about $60\%$. These uncertainties are realistic, and add to those obtained in Table~\ref{Tab1} from the use of different $q_{max}$ around 850 MeV, which are of similar order, and even bigger. Their global consideration allows us to safely conclude that the tensor exchange is completely negligible.

\begin{figure}[H]
\centering
\subfigure[]
{\includegraphics[width=0.328\textwidth]{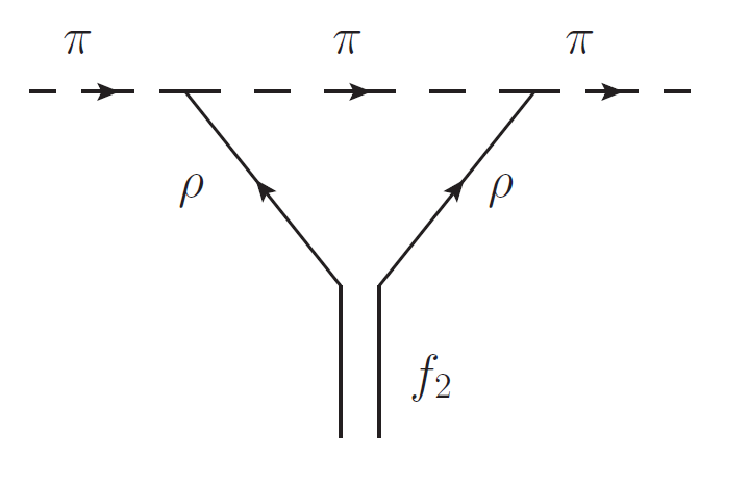}}
\hspace{1.5cm} ~~~
\subfigure[]
{\includegraphics[width=0.245\textwidth]{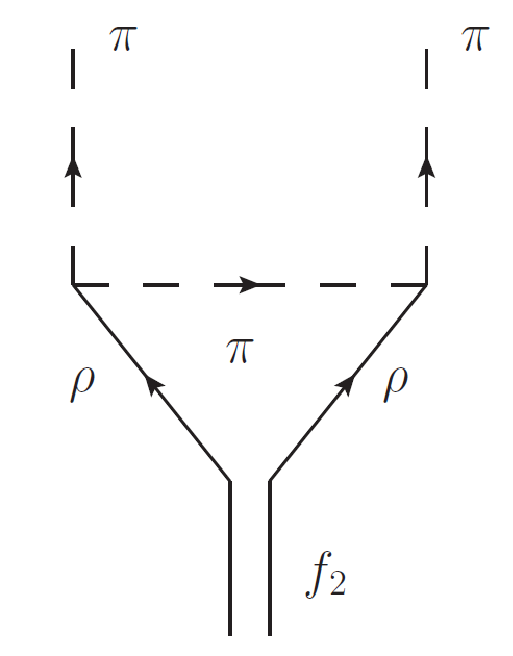}}
\caption{ Diagrams for $f_2(1270)$ coupling to $\pi\pi$. (a) An incoming and an outgoing pion; (b) two outgoing pions. }
\label{newfig}
\end{figure}

Starting from this point let us see what are the difference between an approach like the one in~\cite{Ecker:2007us} and the present one.
In~\cite{Ecker:2007us} a coupling of $f_2(1270)$ to  $\pi\pi$ is assumed and taken from the $f_2(1270)\to \pi\pi$ width. Here we do not use this coupling but we see that our picture produces the right $ \pi\pi$ decay. Now let us compare the diagrams  (a) and (b) of Fig.~\ref{newfig}.
In Fig.~\ref{newfig}(a) the $f_2(1270)$ carries zero energy at the $\pi$ threshold. Hence the two $\rho$  mesons together carry zero energy and {hence} are very off shell, to the point that the whole vertex could be considered a contact term, as implicitly assumed in~\cite{Ecker:2007us}. However, in the diagram of Fig.~\ref{newfig}(b), for the decay of the $f_2(1270)$,  the two $\rho$ mesons {carry} 1270~MeV and
they are almost on shell. Hence, it is clear that the loop function in Fig.~\ref{newfig}(b) is  bigger than that in Fig.~\ref{newfig}(a). If one assumes  the same coupling to $\pi\pi$ in both processes, as implicitly assumed in~\cite{Ecker:2007us}, the large reduction of the loop of  Fig.~\ref{newfig}(a) with respect to the one of  Fig.~\ref{newfig}(b)
is lost. This is why we obtain {about} a factor $2.6$ smaller result for the tensor exchange in our picture than in Ref.~~\cite{Ecker:2007us} for $\pi$ scattering, while having the same decay width  to $\pi\pi$.

Next, we try to understand why the ratio for $K$ scattering is about a factor 4.6 times larger than in the case of the pion while in the case of~\cite{Ecker:2007us} it is a factor of about 17 times larger. This factor 17 in~\cite{Ecker:2007us} is easy to understand. Assuming SU$(3)$ symmetry,
as in~\cite{Ecker:2007us}, one expects roughly that the ratio $R$  for pions or Kaons, 
goes to $ R_K/R_\pi \approx m_K^2/m^2_\pi \approx 13 $, 
not far from the factor 17 obtained in the detailed  calculation. However, in the present case, due to the loops involved in the use of the dynamically generated $f_2(1270)$, 
we get an extra factor $[(\frac{8}{3})^2\frac{1}{4}]^{-1}=0.56$ reduction (see $C_j$ coefficients after Eq.~(\ref{vertex5})) in the case of Kaons compared to the case of pions. This factor  times 13 gives 7.3 to be compared with the factor 4.6 that our calculation gives. The extra reduction should be associated to the structure of $V_{00}$ in Eq.~(\ref{vertex5}) with has powers of $w_i$ in the denominator. We see, then, that the mechanism of tensor exchange, assuming the $f_2(1270)$ dynamically generated, has  a visible effect, reducing appreciably what one finds with the use of SU$(3)$ symmetry in~\cite{Ecker:2007us}.

For $D$ scattering the ratio $R$ is bigger, of the order of $1.7 \%$ but still a small fraction that  also makes this contribution negligible. The structure of the loop faction $V_{00}$ of {Eq.~(\ref{vertex5})} with the large mass of the $D$ meson stabilizes the results.

In the end, in the studies of elastic processes involving pseudoscalar mesons, for practical purposes one can take into account only the contributions coming from the vector-exchange mechanism, and neglect the tensor-exchange mechanism.

\begin{table}[H]
\footnotesize
\centering
 \caption{Ratio $R$ defined in Eq.~(\ref{ratio}) for the processes $\pi^+ \pi^-\to \pi^+ \pi^-$, $K^+ K^-\to K^+ K^-$ and $D^+ D^-\to D^+ D^-$ taking different values for the coupling $g_{f_2}$ reported in Refs.~\cite{Geng2008,Molina:2008jw}, and for the cutoff $q_\mathrm{max}$.}
 \label{value_TV}
\setlength{\tabcolsep}{22pt}
\begin{tabular}{ccccc}
\hline
\multirow{1}{*}{$g_{f_2}$[MeV]} &$q_\mathrm{max}$[MeV] &  $\pi^+ \pi^-\to \pi^+ \pi^-$  &  $K^+ K^-\to K^+ K^-$  &$D^+ D^-\to D^+ D^-$ \\
\hline
\\
$10551$~\cite{Molina:2008jw} & $600$ & $8.38\times10^{-5} $ & $ 2.960\times10^{-4} $ & $  4.79\times10^{-4}$
\\
& $700$ & $ 3.22\times10^{-4}  $ & $  1.29\times10^{-3} $ & $  2.47\times10^{-3}$
\\
 & $800$ &$ 9.56\times10^{-4}  $ & $  4.22\times10^{-3} $ & $  9.35\times10^{-3}$
\\
 & $850$ &$ 1.53\times10^{-3}  $ & $  7.00\times10^{-3} $ & $  1.66\times10^{-2}$
\\
& $900$ &  $2.34\times10^{-3}  $ & $  1.11\times10^{-2} $ & $  2.81\times10^{-2}$
\\
\\
\hline
\hline
\\
$ 10889$~\cite{Geng2008} & $600$ &  $8.93\times10^{-5} $ & $  3.15\times10^{-4} $ & $  5.10\times10^{-4}$
\\
& $700$ &  $3.43\times10^{-4} $ & $  1.38\times10^{-3} $ & $  2.63\times10^{-3}$
\\
& $800$ &  $1.02\times10^{-3} $ & $  4.49\times10^{-3}  $ & $ 9.96\times10^{-3}$
\\
& $850$ &  $1.63\times10^{-3} $ & $  7.45\times10^{-3}  $ & $ 1.77\times10^{-2}$
\\
& $900$ &  $2.49\times10^{-3} $ & $  1.18\times10^{-2} $ & $  2.99\times10^{-2}$
\\
\\
\hline  
\hline
   \end{tabular}
\label{Tab1}   
\end{table}

\section{Conclusions}

In this work we have studied the tensor and vector-contribution mechanisms to the elastic reactions involving the pseudoscalar mesons $\pi^+ \pi^-$, $K^{+}K^{-}$ and $D^{+}D^{-}$. In the case of the tensor-exchange contribution we have assumed that an intermediate tensor $f_2(1270)$ is dynamically generated from the $\rho\rho$ interaction with the use of a pole approximation, which makes the two-loop calculation easier since the triangle loops can be factorized and computed separately. The results have shown small contributions coming from the tensor-exchange mechanisms when compared with those from the vector-exchange processes. Hence, for pragmatic considerations only the vector-exchange mechanisms are relevant, and the tensor-exchange contributions can be neglected. We have also compared our results with those of~\cite{Ecker:2007us} and shown the similarities in the input from both pictures for pion scattering, where the results are more ressemblant. Yet, we showed that while the SU$(3)$ picture assumed in~\cite{Ecker:2007us} provides a relative tensor contribution for $K\bar{K}$ scattering about $17$ times larger than the one for pions, the picture where the $f_2(1270)$ is dynamically generated stabilizes the tensor contribution, which is then only about 4.6 times bigger  than the one of the pion. Our picture also allows us to calculate the ratio of tensor to vector exchange for the case of $D$ mesons that we find of the order of $1-2 \%$.

\section{acknowledgements}

The work of P.C.S.B and L.M.A. is partly supported by the Brazilian agencies CNPq (Grant Numbers 309950/2020-1, 400215/2022-
5, 200567/2022-5), FAPESB (Grant Number INT0007/2016) and CNPq/FAPERJ under the Project INCT-F\'{\i}sica Nuclear e
Aplicações (Contract No. 464898/2014-5).
This work of J. S. is partly supported by the National Natural Science Foundation of China under Grants No. 12247108 and the China Postdoctoral Science Foundation under Grant No. 2022M720359. 
This work is
also partly supported by the Spanish Ministerio de Economia y Competitividad (MINECO) and European FEDER
funds under Contracts No. FIS2017-84038-C2-1-P B, PID2020-112777GB-I00, and by Generalitat Valenciana under
contract PROMETEO/2020/023. This project has received funding from the European Union Horizon 2020 research
and innovation programme under the program H2020-INFRAIA-2018-1, grant agreement No. 824093 of the STRONG-2020 project. This research is also supported by the Munich Institute for Astro-, Particle and BioPhysics (MIAPbP)
which is funded by the Deutsche Forschungsgemeinschaft (DFG, German Research Foundation) under Germany’s
Excellence Strategy-EXC-2094 -390783311.


%

\end{document}